\documentclass[a4paper,11pt]{article}
\usepackage{pos}

\def\beq{\begin{equation}}
\def\eeq{\end{equation}}
\def\bea{\begin{eqnarray}}
\def\eea{\end{eqnarray}}

\title{Global analyses of nuclear PDFs with heavy-quark and neutrino data}
\author*[a,b]{M.~Klasen}
\author[c,d]{P.~Duwent\"aster}
\author[a]{T.~Je\v{z}o}
\author[a]{K.~Kova\v{r}\'{\i}k}
\author[e]{A.~Kusina}
\author[f]{J.G.~Morf\'{i}n}
\author[a]{K.F.~Muzakka}
\author[g]{F.I.~Olness}
\author[e]{R.~Ruiz}
\author[h]{I.~Schienbein}
\author[h]{J.Y.~Yu}

\affiliation[a]{Institut f{ü}r Theoretische Physik, Westf{ä}lische Wilhelms-Universit{ä}t
M{ü}nster, Wilhelm-Klemm-Stra{ß}e 9, 48149 M{ü}nster, Germany}
\affiliation[b]{School of Physics, The University of New South Wales, Sydney NSW 2052, Australia}
\affiliation[c]{University of Jyväskylä, Department of Physics, P.O.\ Box 35, FI-40014 University of Jyväskylä, Finland}
\affiliation[d]{Helsinki Institute of Physics, P.O.\ Box 64, FI-00014 University of Helsinki, Finland}
\affiliation[e]{Institute of Nuclear Physics, Polish Academy of Sciences, ul. Radzikowskiego, Cracow 31-342, Poland}
\affiliation[f]{Fermi National Accelerator Laboratory, Batavia, IL 60510, USA}
\affiliation[g]{Southern Methodist University, Dallas, TX 75275, USA }
\affiliation[h]{Laboratoire de Physique Subatomique et de Cosmologie, Université Grenoble-Alpes, CNRS/IN2P3, 53 avenue des Martyrs, 38026 Grenoble, France}

\emailAdd{michael.klasen@uni-muenster.de}

\abstract{
We discuss the two most recent global analyses of nuclear parton distribution functions within the nCTEQ approach. LHC data on $W/Z$-boson, single-inclusive hadron and heavy quark/quarkonium production are shown to not only significantly reduce the gluon uncertainty down to $x\geq10^{-5}$, but to also influence the strange quark density. The latter is further constrained by neutrino deep-inelastic scattering and charm dimuon production data, whose consistency with neutral-current experiments is also re-evaluated.
}

\FullConference{%
  41st International Conference on High Energy physics - ICHEP2022\\
  6-13 July, 2022\\
  Bologna, Italy
}

\begin{document}
\hfill MS-TP-22-39
\maketitle

%%% Section 1 %%%%%%%%%%%%%%%%%%%%%%%%%%%%%%%%%%%%%%%%%%%%%%%%%%%%%%%%%%%%%%%%%%
\section{Motivation}

The nuclear structure at high energies is not only an interesting research topic by itself, as we seek to understand the fundamental quark and gluon dynamics of protons and neutrons bound in nuclei, but it also represents an important initial condition in the creation of a new state of matter, the quark-gluon plasma. While the evolution of parton density functions (PDFs) $f_{q,g}^{p,n/A}(x,Q)$ with the scale $Q$ is calculable in perturbative QCD through the DGLAP evolution equations, the dependence on the longitudinal momentum fraction $x$ is not. Thanks to the QCD factorization theorem, it is, however, universal and can therefore be determined in global fits to a variety of experimental data. Similarly, the fundamental dynamics of nuclear modifications such as shadowing, antishadowing, the EMC effect and Fermi motion, can be parameterized, but remains to be fully theoretically understood.

Global analyses of nuclear PDFs (nPDFs) have been and are performed by the nCTEQ, DSSZ, EPPS, HKN, KSASG, nNNPDF and TUJU collaborations (cf.\ the papers cited in Ref.\ \cite{Kusina:2020lyz}). Within the nCTEQ approach, the coefficients in the $x$-dependence of the nPDFs,
\bea
 xf_i^{p/A}(x,Q_0)&=&c_0x^{c_1}(1-x)^{c_2}e^{c_3x}(1+e^{c_4}x)^{c_5},\\
 c_k&\to&c_{k,0}+c_{k,1}(1-A^{-c_{k,2}})
\eea
depend directly on the nuclear mass number $A$, and the nPDFs are evolved for each quark flavor and the gluon. Neutron PDFs are obtained from isospin symmetry. The nCTEQ15 fit was based on fixed target data from deep-inelastic scattering (DIS) and Drell-Yan (DY) experiments to constrain the quarks and antiquarks as well as RHIC neutral pion data to constrain the gluon. It was thus limited in its precision, in particular at low $x$ and for the strange quark. LHC data and a careful reanalysis of neutrino DIS and charm dimuon data now offer the possibility to considerably improve on this fit in both directions.

%%% Section 2 %%%%%%%%%%%%%%%%%%%%%%%%%%%%%%%%%%%%%%%%%%%%%%%%%%%%%%%%%%%%%%%%%%
\section{Heavy-quark and quarkonium data}

The four LHC experiments have provided data on $W/Z$ boson and single inclusive hadron production in $pA$ scattering, which have led to improved determinations of nPDFs within the nCTEQ15WZ \cite{Kusina:2020lyz} and nCTEQ15WZSIH \cite{Duwentaster:2021ioo} global fits. While the main impact was on the gluon density, whose uncertainty in the region $x\sim10^{-3}...10^{-2}$ at $Q=2$ GeV was reduced twice by factors of two, also the strange quark density at these values of $x$ was affected and in particular pulled to considerably larger values, although its uncertainty remained substantial.

\begin{figure}\centering
 \includegraphics[width=0.49\textwidth]{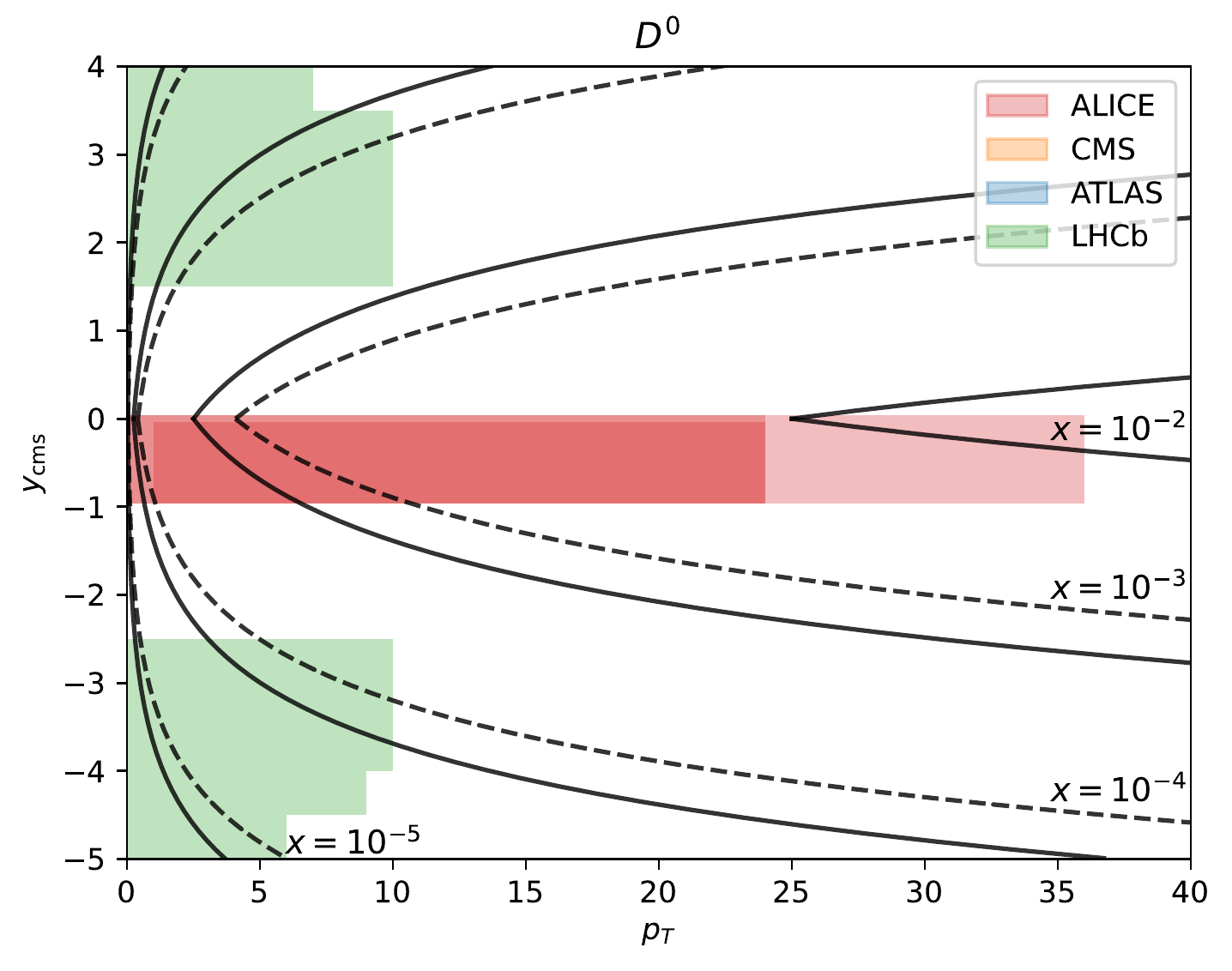}
 \includegraphics[width=0.49\textwidth]{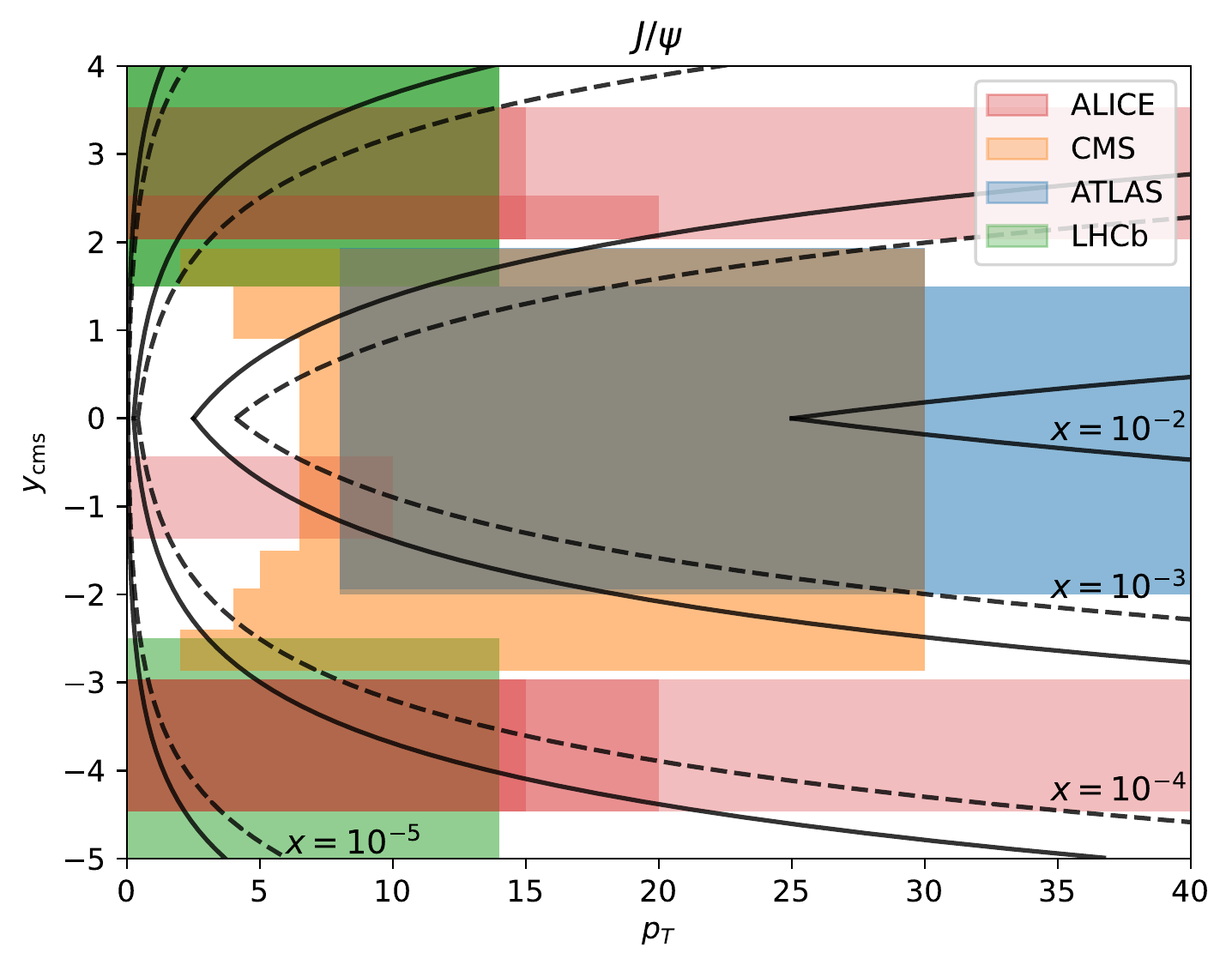}
 \includegraphics[width=0.49\textwidth]{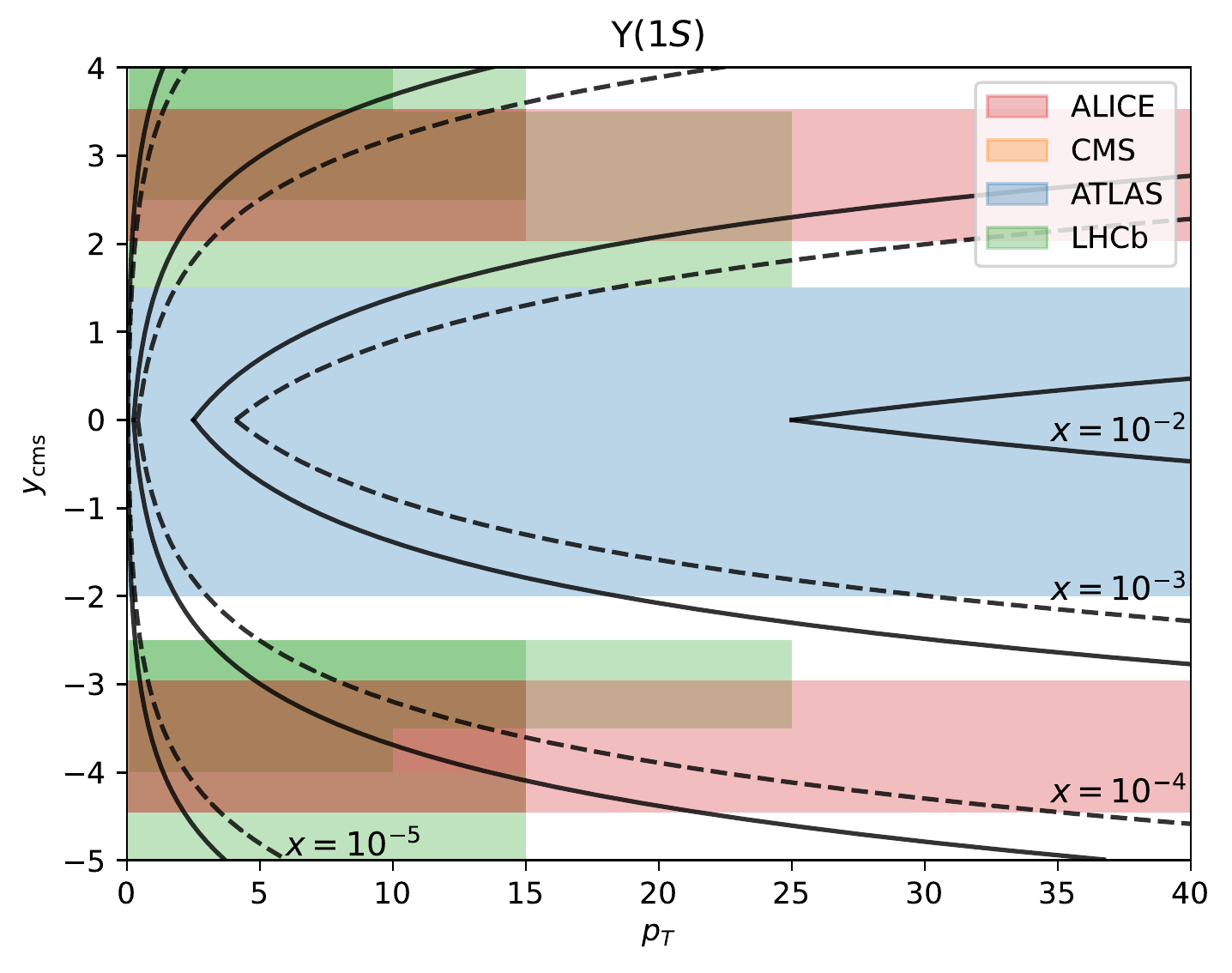}
 \includegraphics[width=0.49\textwidth]{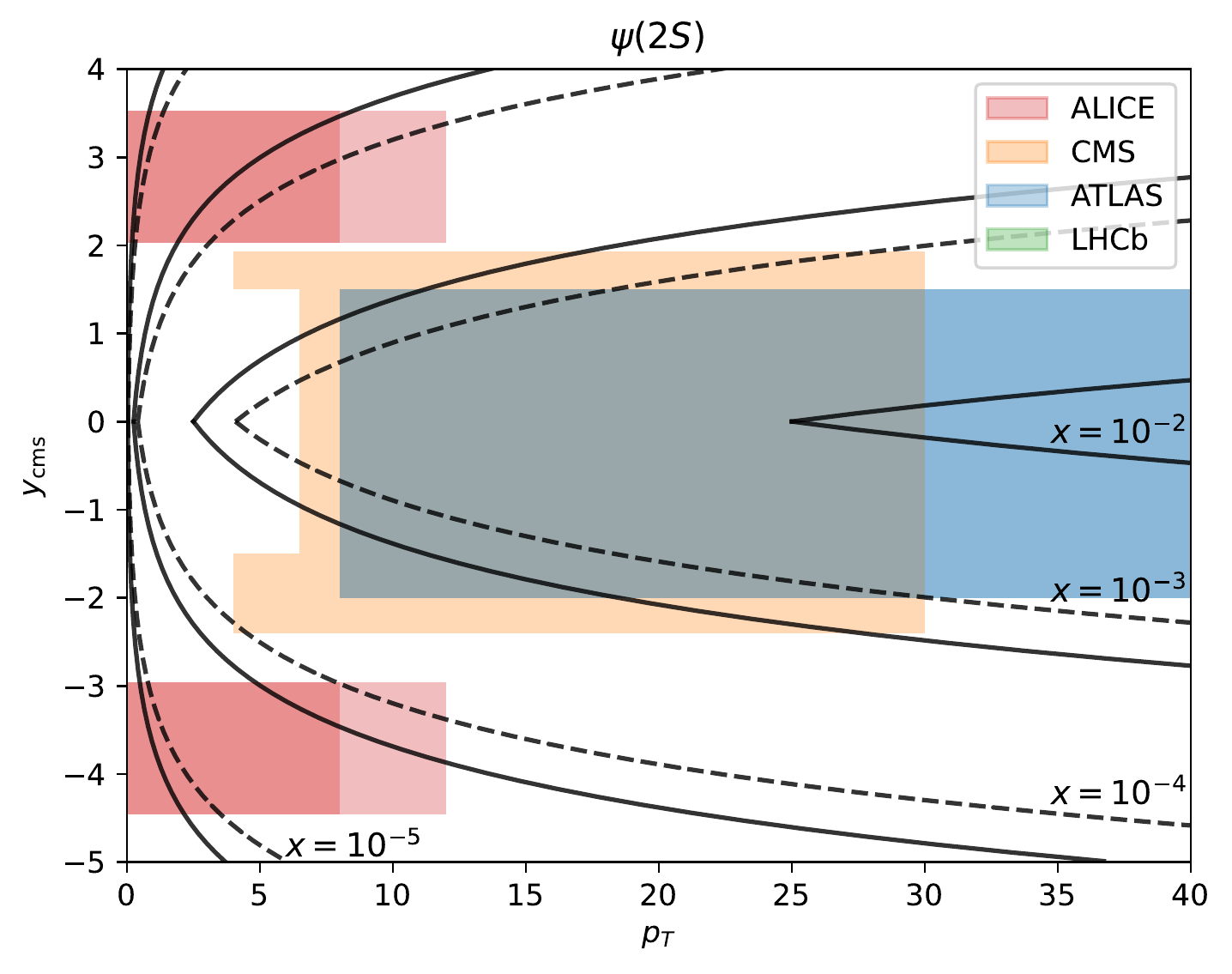}
 \caption{Coverage of the kinematic $(p_T,y_{\mathrm{cms}})$-plane of the quarkonium and open heavy quark production data sets from proton-lead collisions. ALICE data is shown in red, ATLAS in blue, CMS in orange and LHCb in green. The dashed and solid contours show the estimated $x$-dependence for $\sqrt{s}=5$ and 8\,TeV, respectively.}
 \label{fig:1}
\end{figure}

A sizable data set, covering a large region in the kinematic (transverse-momentum $p_T$, center-of-mass system rapidity $y_{\rm cms}$) plane is now available from heavy quark and quarkonium production at the LHC (see Fig.\ \ref{fig:1}). A naive estimate shows that it allows to probe values of $x={2p_T/\sqrt{s}}\exp(-|y|)$ down to $10^{-5}$. This holds the promise for again much improved gluon and sea quark densities at low $x$, which would e.g.\ significantly reduce the dominant error in the predicted charm meson abundances in the statistical hadronization model \cite{Andronic:2017pug}.

A fast evaluation of the heavy quark and quarkonium production cross sections in $pA$ collisions as required in global nPDF fits is possible, when they are assumed to be dominated by gluon scattering, parametrized with a Crystal Ball function and fitted to the well measured cross sections in $pp$ collisions \cite{Kusina:2017gkz}. The large regions of the kinematic plane (accessed in particular by LHCb) require, however, to extend the ansatz to
\begin{align}
\begin{split}
&\overline{\left|\mathcal{A}_{g g \rightarrow \mathcal{Q}+X}\right|^{2}}= \frac{\lambda^2\kappa\hat{s}}{M_\mathcal{Q}^2} e^{a|y_{\rm cms}|}
\times\begin{cases} e ^{ -\kappa \frac{p_{T}^{2}}{M_{\mathcal{Q}}^{2}}} & \text { if } p_{T} \leq\left\langle p_{T}\right\rangle \\ e ^ {-\kappa \frac{\left\langle p_{T}\right\rangle^{2}}{M_{\mathcal{Q}}^{2}}}\left(1+\frac{\kappa}{n} \frac{p_{T}^{2}-\left\langle p_{T}\right\rangle^{2}}{M_{\mathcal{Q}}^{2}}\right)^{-n} & \text { if } p_{T}>\left\langle p_{T}\right\rangle\end{cases} \nonumber
\end{split}
\end{align}
i.e.\ to include a rapidity dependence. The corresponding fits to the $pp$ baseline have been validated with next-to-leading order (NLO) calculations in QCD \cite{Kniehl:2012ti} and NRQCD \cite{Butenschoen:2010rq}.

\begin{figure}\centering
 \includegraphics[width=0.95\textwidth]{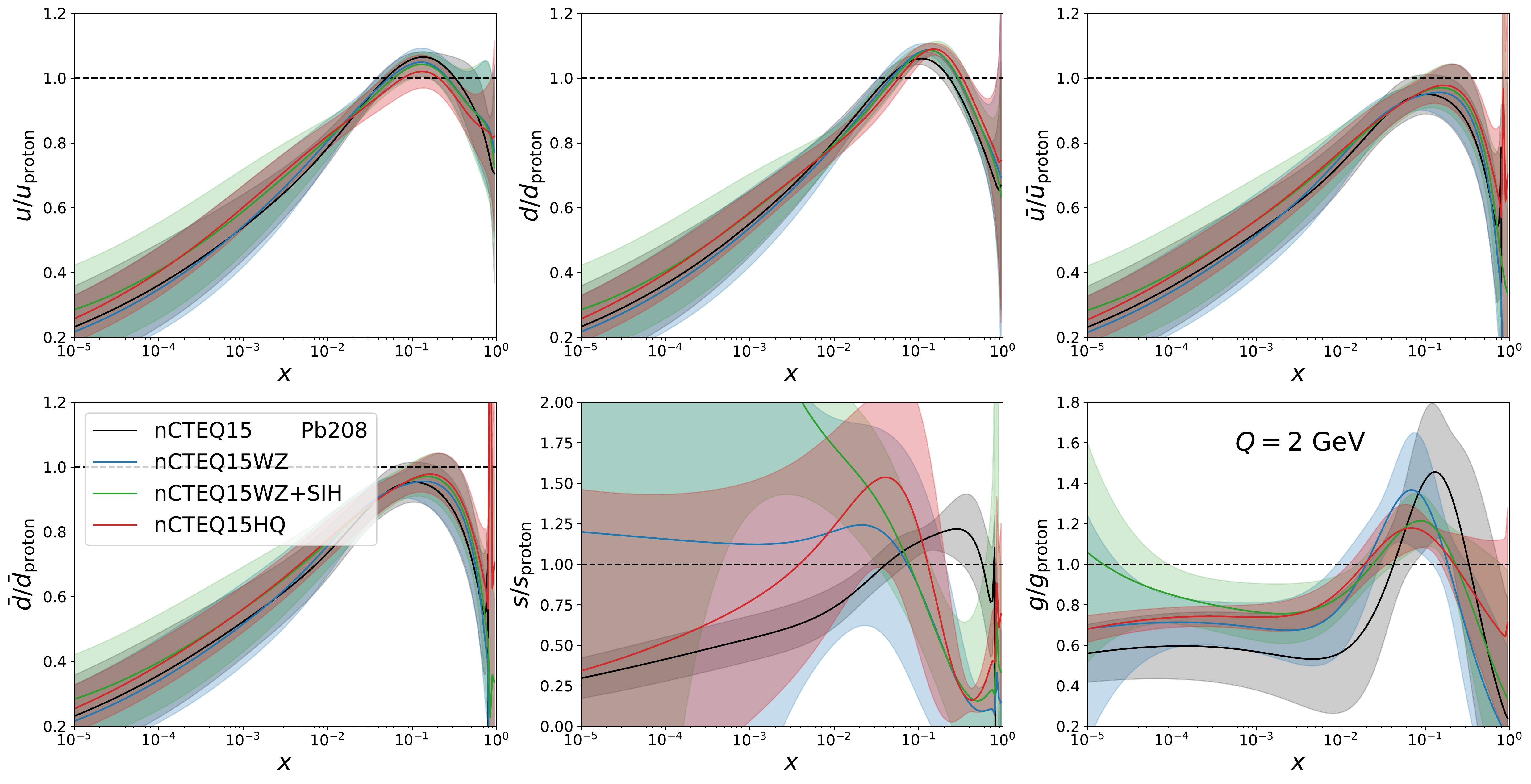}
 \caption{Ratio of lead and proton PDFs from different nCTEQ15 versions. The baseline nCTEQ15 fit is shown in black, nCTEQ15WZ in blue, nCTEQ15WZSIH in green, and the new nCTEQ15HQ fit in red.}
 \label{fig:2}
\end{figure}

The result, shown in Fig.\ \ref{fig:2} (red), is indeed impressive, in particular when compared to the previous nCTEQ15 (black), nCTEQ15WZ (blue) and nCTEQ15WZSIH (green) fits. The value of $\chi^2$ per degree of freedom (dof) decreases from 1.23 to 0.90, 0.92 and 0.86, respectively, despite the fact that the latter three fits have been performed with 19 instead of the 16 free parameters of the original nCTEQ15 fit. This is of course due to the addition of (precise and compatible) data points, which increases their number from 740 in nCTEQ15 to 860 in nCTEQ15WZ, 948 in nCTEQ15WZSIH and 1484 in nCTEQ15HQ. The gluon density is now determined down to $x=10^{-5}$ with an uncertainty that is almost an order of magnitude smaller than in nCTEQ15WZSIH. Interestingly, the rise in the strange quark density observed there is not confirmed in the new fit, although its uncertainty remains again large \cite{Duwentaster:2022kpv}.

%

%%% Section 3 %%%%%%%%%%%%%%%%%%%%%%%%%%%%%%%%%%%%%%%%%%%%%%%%%%%%%%%%%%%%%%%%%%
\section{Neutrino DIS and dimuon data}

Charged current (CC) neutrino DIS and charm dimuon production data have long been known to be very sensitive to the strange quark density. The compatibility of the corresponding measurements from CDHSW, CCFR, NuTeV and Chorus with neutral current (NC) DIS measurements has, however, been critically discussed in the literature \cite{Kovarik:2010uv,Paukkunen:2013grz}.

We have re-evaluated the tensions of the neutrino data both internally and with the data sets used in nCTEQ15WZSIH, taking into account nuclear effects in the calculation of the deuteron structure function with a new baseline fit (nCTEQ15WZSIHdeut) \cite{Segarra:2020gtj}. Increasing the number of data points from 740 in nCTEQ15 to 940 required an increase in the tolerance from $\Delta\chi^2=35$ to 45. We found that even with this improved baseline and larger tolerance the tensions persist, independently of the proton baseline and the treatment of data correlation and normalization uncertainties. The tension can just be relieved using a kinematic cut of $x > 0.1$ with an increase of $\Delta\chi^2$ by 46 for 4644 neutrino data points, but a global fit using only Chorus and dimuon data (BaseDimuChorus) leads to a much better fit without any additional cuts and an overall increase of $\Delta\chi^2$ by 5 for 974 neutrino data points.

\begin{figure}\centering
 \includegraphics[width=0.95\textwidth]{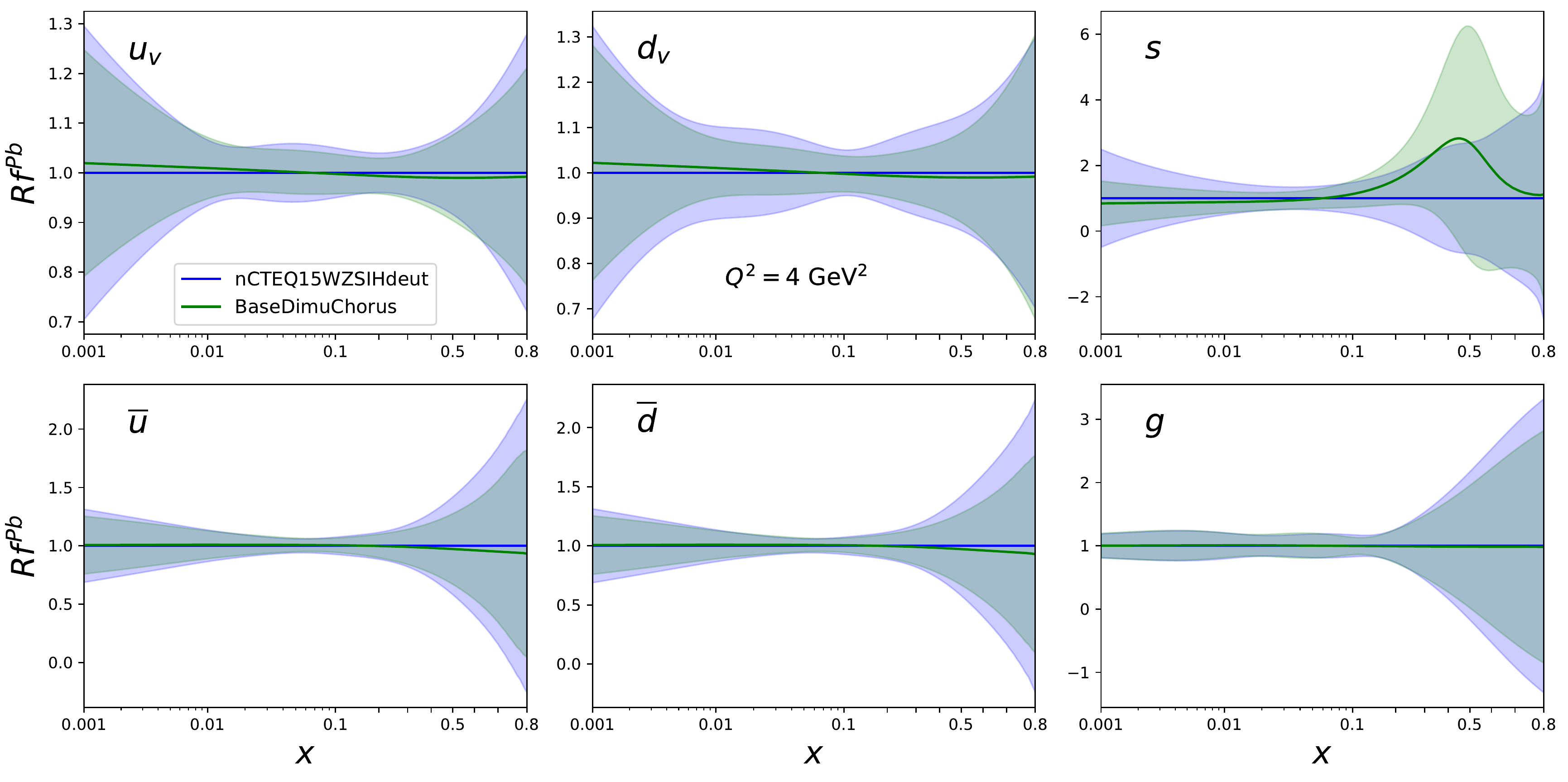}
 \caption{The fitted lead PDF ratio to nCTEQ15WZSIHdeut. All uncertainty bands are obtained using the Hessian method with $\Delta \chi^2=45$.}
 \label{fig:3}
\end{figure}

In Fig. \ref{fig:3} we compare the PDFs from our best neutrino fit (BaseDimuChorus) to those from the baseline fit (nCTEQ15WZSIHdeut). As one can see, the neutrino data affect only the central value of the strange quark and at the same time considerably reduce its uncertainty. In addition, the Chrous (anti-)neutrino cross section data lead to reduced uncertainties of the valence quark PDFs. Breaking down the contributions to the total $\chi^2$ from different classes of experiments as a function of the free parameters shows that the valence quark and the antiquark parameters are still mainly constrained by the NC DIS experiments and the gluon parameters by $W/Z$ and SIH production at the LHC, while the strange quark parameters are constrained by the Chorus and dimuon data \cite{Muzakka:2022wey}.

%

%%% Section 4 %%%%%%%%%%%%%%%%%%%%%%%%%%%%%%%%%%%%%%%%%%%%%%%%%%%%%%%%%%%%%%%%%%
\section{Conclusions and outlook}

In conclusion, the LHC has produced a wealth of precise data on heavy quark and quarkonium production in $pA$ collisions covering large regions of the kinematic plane in transverse momentum and rapidity, that provide access to very low values of the parton momentum fraction in heavy nuclei. These data are well described by a data-driven approach, which is both in agreement with NLO QCD and NRQCD calculations and allows for a sufficiently fast evaluation of the cross section in global fits of nPDFs. Consequently, our new fit nCTEQ15HQ that includes these data constrains the gluon density down to $x=10^{-5}$ with an uncertainty that is almost an order of magnitude smaller than in previous global fits such as nCTEQ15.

There exists also a wealth of high statistics, but older fixed-target neutrino DIS and charm dimuon production data, which have a large potential for flavor separation and are therefore even regularly used in proton PDF analyses. We have re-assessed the compatibility of these data with NC DIS and other data with a deuteron-improved baseline using three different criteria, finding that the tensions persisted independently of the proton baseline and the treatment of data correlation and normalization uncertainties. A kinematic cut of $x>0.1$ can just relieve the tensions, but using only Chorus and dimuon data leads to a much better fit, better constrains the strange quark and also reduces the uncertainties on the valence quarks and the antiquarks.

Due to lack of space, we have not discussed here our recent study of JLab DIS data at high values of $x$ and intermediate-to-low values of $Q$ \cite{Segarra:2020gtj} as well as ongoing work on other LHC data, which will soon lead to a new release of global nPDFs within the nCTEQ approach.

\acknowledgments

M.K.\ thanks the organizers of ICHEP 2022 for the invitation to this very nice conference and the School of Physics at the University of New South Wales in Sydney, Australia for its hospitality as well as for financial support through the Gordon Godfrey visitors program. The work of P.D., T.J., M.K., K.K.\ and K.F.M.\ was funded by the Deutsche Forschungsgemeinschaft (DFG) through projects GRK 2149, SFB 1225, and KL 1266/10-1. This research was also funded as a part of the Center of Excellence in Quark Matter of the Academy of Finland (project 346326). This manuscript has been authored by Fermi Research Alliance, LLC under Contract No.\ DEAC02-07CH11359 with the U.S. Department of Energy, Office of Science, Office of High Energy Physics. F.I.O. was supported by the U.S. Department of Energy Grant No. DE-SC0010129. A.K. and R.R. acknowledge the support of Narodowe Centrum Nauki under Sonata Bis Grant No. 2019/34/E/ST2/00186. R.R. acknowledges the support of the Polska Akademia Nauk (grant agreement PAN.BFD.S.BDN. 613. 022. 2021 - PASIFIC 1, POPSICLE). This work has received funding from the European Union’s Horizon 2020 research and innovation program under the Skłodowska- Curie grant agreement No. 847639 and from the Polish Ministry of Education and Science. The work of I.S. was supported in part by the French National Centre for Scientific Research CNRS through IN2P3 Project GLUE@NLO.

%%% Bibliography %%%%%%%%%%%%%%%%%%%%%%%%%%%%%%%%%%%%%%%%%%%%%%%%%%%%%%%%%%%%%%%

\end{document}